\documentclass[manuscript]{aastex}

\newcommand{\hst}{{\sl HST\/}}
\newcommand{\egg}{\object{Egg Nebula}}

\slugcomment{}

\shorttitle{Differential Proper-Motion Measurements of The Cygnus Egg Nebula}
\shortauthors{Ueta, Tomasino, and Ferguson}

\begin{document}

\title{Differential Proper-Motion Measurements of The Cygnus Egg Nebula:
The Presence of Equatorial Outflows}

\author{Toshiya Ueta
and
Rachael L.\ Tomasino}
\affil{Department of Physics and Astronomy, MS 6900, 
University of Denver, Denver, CO 80208}
 
\and

\author{Brian A.\ Ferguson\altaffilmark{1}}
\affil{Space Telescope Science Institute, Baltimore, MD 21218}

\altaffiltext{1}{Currently: Des Moines University Medical School}

\begin{abstract}
 We present the results of differential proper-motion analyses of the
 Egg Nebula (RAFGL 2688, V1610 Cyg) based on the archived two-epoch
 optical data taken with the Hubble Space Telescope. 
 First, we determined that the polarization characteristics of
 the {\egg} is influenced by the higher optical depth of the central
 regions of the nebula (i.e., the ``dustsphere'' of $\sim 10^3$ AU radius),
 causing the nebula illuminated in two steps  
 -- the direct starlight is first channeled into bipolar cavities and
 then scattered off to the rest of the nebula. 
 We then measured the amount of motion of local structures and the signature
 concentric arcs by determining their relative shifts over the 7.25-yr interval. 
 Based on our analysis, which does not rely on the single-scattering
 assumption,
 we concluded that 
 the lobes have been excavated by a linear
 expansion along the bipolar axis for the past $\sim 400$~yr, while 
 the concentric arcs have been generated continuously and moving out
 radially at about 10~km~s$^{-1}$ for the past $\sim 5,500$~yr, and
 there appears to be a colatitudinally-increasing trend 
 in the radial expansion velocity field of the concentric arcs. 
 There exist numerical investigations into the mass-loss modulation by
 the central binary system, which predict such a
 colatitudinally-increasing expansion velocity field in the spiral-shock 
 trails of the mass-loss ejecta.
 Therefore, the Egg Nebula may represent a rare edge-on case of
 the binary-modulated circumstellar environs, corroborating
 the previous theoretical predictions.
\end{abstract}


\section{Introduction}

The \object[Egg Nebula]{Cygnus Egg Nebula} 
(\object[Egg Nebula]{V1610 Cyg}, 
\object[Egg Nebula]{RAFGL 2688}; hereafter, the
\egg) is one of the first infrared sources discovered with an
optical bipolar nebulosity \citep{ney75}. 
This object, usually referred to as a proto-planetary nebula 
(\citealt{kwok93,vanwinckel03}), consists of the central star in 
the post-asymptotic giant branch (post-AGB) phase and the 
physically-detached circumstellar shell, which is a consequence of mass 
loss during the preceding AGB phase.  
While AGB mass loss is qualitatively understood dust-driven, 
the exact physical mechanisms of the structure formation in the
circumstellar shell are still not understood completely. 

The {\egg} is known for its signature bipolar lobes and rather circular
concentric arcs \citep{sahai98}. 
While there are about a dozen sources with such arcs, the
{\egg} is known to possess the largest number of arcs ($> 20$; 
\citealt{hrivnak01,su04}). 
These concentric arcs are qualitatively perceived as manifestations
of some kind of mass loss modulations that took place during the early
AGB phase.  
However, the deduced temporal intervals do not match with
any of the known theoretical temporal intervals (e.g.,
thermal pulsing and surface pulsations of AGB stars;
\citealt{su04}).

One of the direct methods to understand the dynamics of the
circumstellar structure formation is to perform a differential 
proper-motion analysis and {\sl measure} the amount of relative 
translational motion of specific shell structures.
We performed such an analysis for the bipolar lobes and the equatorial
structures of the {\egg} using NICMOS imaging-polarimetry data taken
with a 5.5-yr baseline and found that 
(1) the lobes expanded linearly (i.e., the expansion velocity linearly 
proportional to the distance to the expansion center),
(2) the distance to the {\egg} was $420 \pm 60$~pc, and
(3) the object (star and nebula) itself experienced its own proper
motion at ($13.7 \pm 2.0$, $10.2 \pm 2.0$) mas yr$^{-1}$ 
\citep{ueta06,ueta12}.
However, we were unable to measure differential proper-motion of the
concentric arcs because the NICMOS field of view was too small and the
surface brightnesses were too weak is in the near-IR.

\citet{balick12} performed the latest attempt using 
ACS/HRC and WFC3/UVIS images with a 6.65-yr baseline.
They reported 
(1) a uniform expansion of the bipolar lobes by 2.5\% and 
(2) radial motion of the arcs by 0\farcs07.
However, their analysis was limited to the central 
$\sim 20^{\prime\prime}$ and lacked background reference sources to 
properly align multi-epoch images.
Hence, they were unable to confirm the bulk motion of the nebula and
{\sl assumed} that the shell structures expanded symmetrically.

In this work, we performed differential
proper-motion measurements of the concentric arcs
using the archived imaging-polarimetry data from WFPC2 and ACS/WFC, 
exploiting the data set's largest field of view, longest exposure time,
and longest temporal baseline to overcome the deficiencies of the
previous analyses. 
Below, we describe the data set and reduction procedure (\S 2), 
present the reduced imaging polarimetry data, differential
proper-motion measurements, and results of our analysis (\S 3), 
and summarize our finding (\S 4). 

\section{Observations and Data Reduction}

We used the archived data of the {\egg}
in the F606W band with polarizers taken with WFPC2 (GTO/WF2-6221; PI:
J.\ Trauger) on 1995 July 17 (epoch 1, hereafter) and ACS/WFC
(CAL/ACS-9586; PI: W.\ Sparks) on 2002 October 16 (epoch 2, hereafter),
which provide a 7.25-yr baseline (Table \ref{obs}). 
We reduced the data using the PyRAF/STSDAS MultiDrizzle package version
3.13 \citep{md}\footnote{PyRAF and STSDAS are products of the STScI, which is
operated by AURA for NASA.} to process the 
multi-program data into a common frame of reference.
After removing cosmic-ray hits from the pipeline-calibrated data
using {\sl L.A.Cosmic\/} (\citealt{vd01}; there was only one exposure
with each of the three WFPC2 polarizers), we performed two rounds of
MultiDrizzle processes to refine the image alignment by using background
point sources as spatial anchors.

In the end, we generated two sets of three polarizer images 
for a $60^{\prime\prime} \times60^{\prime\prime}$ field 
at the $0\farcs05$ pix$^{-1}$ scale 
centered at the catalog coordinates of the {\egg}   
($21^{\rm h}02^{\rm m}18\fs27$,
$+36^{\circ}41^{\prime}37\farcs0$; \citealt{sahai98b}).
Finally, these polarizer images were combined into two sets of Stokes
images via matrix transformations for each of the WFPC2 and ACS
polarizers as described by \citet{bm97} and \citet{gonzaga11},
respectively.

We measured the amount of shift for each background source
and rejected those that shifted more than two-$\sigma$ of the
distribution: after this exercise 31 sources were adopted as 
stationary spatial anchors.
Because the {\egg} was imaged near the center of the instrument arrays 
(central $80^{\prime\prime} \times 80^{\prime\prime}$ for WFPC2/WF
and $200^{\prime\prime} \times 200^{\prime\prime}$ for ACS/WFC)
these anchors are not affected by the geometric distortion, which may
contribute to false proper motions near the aperture edges. 
The accuracy of alignment was assessed by a residual vector
root-mean-square offset computed from the Gaussian centroid positions
of these 31 spatial anchors and determined to be as good as 0.36 pix
(0\farcs018).  

\section{Data Analysis and Discussion}

\subsection{Photometric Variability}

The Stokes {\sl I\/} images (Figs.\ \ref{figimages}a,b) 
are the total flux images at the Broad V band in the AB
magnitude system \citep{oke74,gonzaga11}.
These maps show the signature bipolar lobes plus the searchlight beams 
as well as the concentric arcs.
The two-epoch aperture photometry revealed that the total integrated
flux of the {\egg} decreased by $36\%$ over 7.25~yrs,
while the individual fluxes of the N and S lobes decreased by 34\% and
41\%, respectively (Table \ref{photom}). 

\citet{balick12} reported similar but smaller flux variations, with the
N lobe faded more than the S lobe (by $11\%$ and $3\%$,
respectively), over 6.65-yr baseline immediately following ours.
Moreover, \citet{hrivnak10} discovered that the integrated flux of the N
lobe varies with a roughly 90-day period.
These observations suggest that both N and S lobes vary their fluxes
periodically with a phase lag, which is probably a manifestation of the
pulsation of the central star modulated by the time-delay due to the
line-of-sight path-length difference between the two lobes of the
inclined nebula.

\subsection{Polarization Characteristics}

\subsubsection{Polarized-Flux-Only Maps}

\begin{figure}
\begin{center}
 \includegraphics[width=\textwidth]{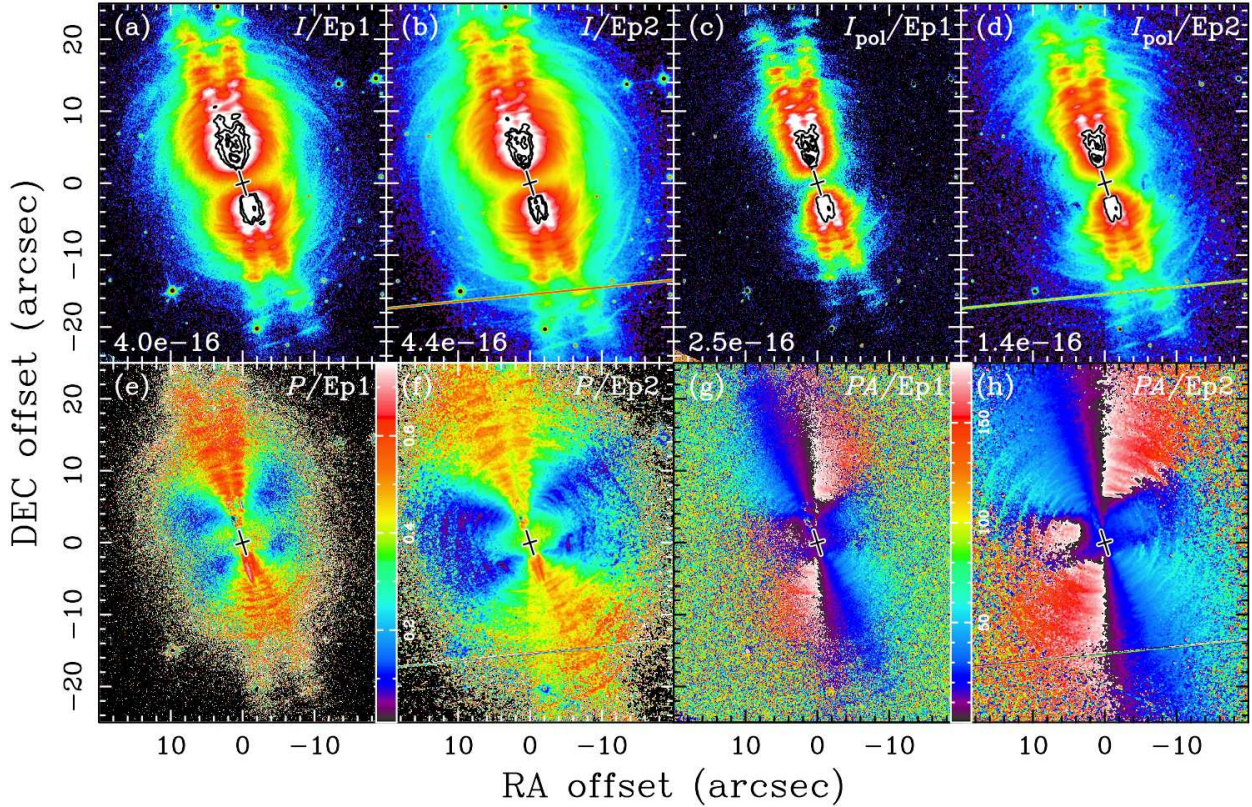} 
\end{center}
\caption{Two-epoch F606W images of the Egg Nebula in the total
 intensity, $I$ (a, b), linear-polarization-flux-only, $I_{\rm pol}$ 
 (c, d), polarization strength, $P$ (e, f), and polarization angle,
 {\sl PA} (g, h).  
 Each pair displays both epoch 1 (Ep1, left) and epoch
 2 (Ep 2, right) images. 
 The $I$ and $I_{\rm pol}$ maps show surface brightnesses above 1 $\sigma$
 ($4.3, 1.0, 6.1$, and $0.8 \times 10^{-20}$ erg cm$^{-2}$ s$^{-1}$
 \AA$^{-1}$ pix$^{-1}$, respectively for (a) to (d), in black) up to 
 $1.0 \times 10^{-17}$ erg cm$^{-2}$ s$^{-1}$ \AA$^{-1}$ pix$^{-1}$ (in white).
 Contours indicate 90, 50, 10, and 1\% levels of the peak surface
 brightness, which is shown at lower-left in 
 erg cm$^{-2}$ s$^{-1}$ \AA$^{-1}$ pix$^{-1}$.
 The $P$ maps show polarization strengths in fraction as indicated in
 the color wedge, while the {\sl PA} maps exhibit polarization vector
 angles in degrees measured from W/$+x$ as indicated in the color
 wedge. 
 All maps are centered at the position of the illumination source
 derived in our previous NICMOS differential proper-motion analysis
 \citep{ueta06}, at which the plus symbol is to indicate
 the one-$\sigma$ error ellipse.\label{figimages}}
\end{figure}

The linearly-polarized-flux-only images,
$I_{\rm pol}$, were constructed with 
the Stokes {\sl Q\/} and {\sl U\/} images 
($I_{\rm pol} = \sqrt{Q^2 + U^2}$; Figs.\ \ref{figimages}c,d).   
In the single-scattering limit, $I_{\rm pol}$ images represent 
the cross-sections of the circumstellar shell because 
$I_{\rm pol}$ tends to be the strongest when the angle of scattering is
close to $90^{\circ}$.
This uniqueness of $I_{\rm pol}$ had been used previously to probe 
the dust density structure in the evolve star circumstellar shells
\citep{ueta05,ueta07}.
 
Contrary to our expectations, however, the bipolar lobes and searchlight 
beams appeared more prominently than the concentric arcs 
(Figs.\ \ref{figimages}c,d).
This indicates that the nebula cannot be approximated in the
single-scattering limit (as evidenced by the presence of the dust lane):
the {\egg} is therefore illuminated by a two-step process, in which 
starlight is first directed into the lobes 
and then scattered off from the lobes to the rest of the nebula.

\subsubsection{Polarization Strength Maps}

The polarization strength images, $P = I_{\rm pol}/I$, 
display the relative strength of polarization more clearly
 (Figs.\ \ref{figimages}e,f).
While stronger polarizations ($\gtrsim 40\%$) are seen in the lobes and 
searchlight beams, weaker polarizations ($\lesssim 40\%$) are
dominant along the equatorial plane except for the region of medium
polarizations (about 30 to $50\%$, delineating the 
central $2.4^{\prime\prime} \times 3.8^{\prime\prime}$ region 
around the dust lane 
($1.0 \times 10^3$ by $1.6 \times 10^3$ AU at 420 pc).

This medium-polarization region appears to represent
the surface of a marginal central dust concentration 
(often referred to as a dust cocoon;
\citealt{latter93,sahai98b,sahai98,goto02})
at which the line-of-sight optical depth at $V$ becomes greater than
unity (i.e., dustsphere as in photosphere and MOLsphere).
This dustsphere and the bipolar lobes are essentially brightly-lit
surfaces that illuminate the rest of the nebula.
Hence, dust-scattered light from the rest of the
nebula tends to be de-polarized, i.e., weak in $I_{\rm pol}$
and $P$. 

The structure of this central dust concentration is still unknown.
\citet{cox00} reported the presence of a hollow CO shell of
$1^{\prime\prime}$ radius expanding at $\sim 10$~km~s$^{-1}$ at the
position angle of $\sim 54^{\circ}$ (coincident with the orientation of
the 1.3 mm dust continuum). 
Thus, this medium-polarization-strength region
could represent the ``near-side surface'' of this central expanding  
CO/dust shell. 
To better assess the geometry of the central material distribution,
further investigations at higher spatial resolution at optically-thin
bands are required. 

\subsubsection{Polarization Angle Maps}

The polarization angle ({\sl PA}~$= \frac{1}{2}\arctan(U/Q)$) images
show the position angle of the 
electric vector of the incident light (Figs.~\ref{figimages}g,h).
The {\sl PA} images for optically-thin nebulae typically show a pattern
of an azimuthal gradient, which represents continuously changing
polarization vector angles with respect to the position of the
illumination source. 
However, our maps reveal roughly the same angle near the center
generally aligned with the dust lane ($107^{\circ}$ of E of N),
indicating higher dust density around the dust lane at which the
single-scattering approximation breaks down.
For this reason, we were unable to pinpoint the location of the
illumination source by back-tracing the polarization vectors 
(e.g., \citealt{wk93,weintraub00,ueta06})
more accurately than a one-$\sigma$ error ellipse of 
$3.4^{\prime\prime} \times 1.8^{\prime\prime}$.

\subsection{Differential Proper-Motion Measurements}

\subsubsection{Nebula Expansion}

As discussed above, the optical reflection nebulosity of the {\egg} is
caused by a two-step illumination process in rather de-polarized light.
Therefore, we reverted to the total intensity $I$ maps (i.e., to utilise
all the flux available to us) to follow
the differential proper-motion of local shell structures using the same 
correlational method as in our previous analysis.
Briefly, this method measures the amount of translational shift of a
given structure between two epochs via a cross-correlation analysis
between cutouts of the two-epoch images (e.g., \citealt{currie96,morse01}).
We use our own IDL script set created as part of our previous
proper-motion investigations of the {\egg} in the near-IR
(\citealt{ueta06} for details). 

\begin{figure}
\begin{center}
\includegraphics[width=\textwidth]{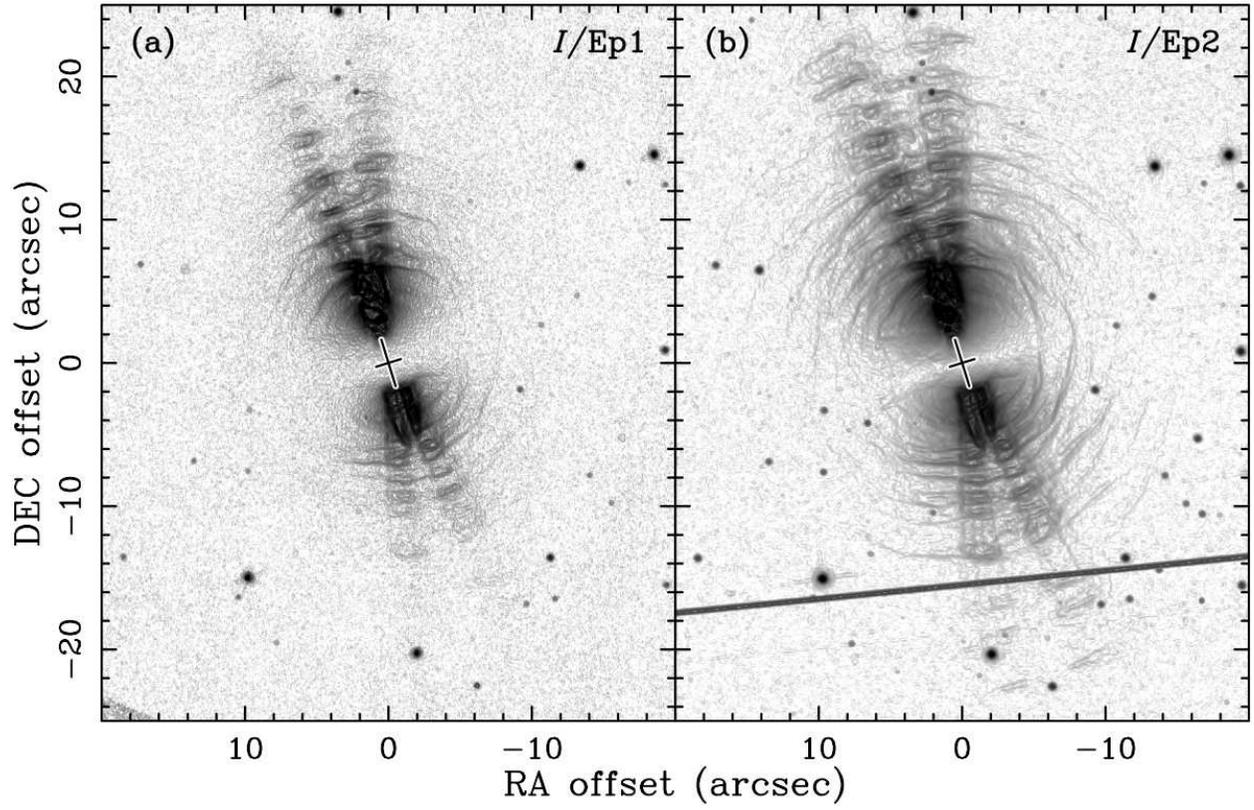} 
\end{center}
\caption{The edge-enhanced Stokes $I$ maps of the
 Egg Nebula: (epoch 1 on left and epoch 2 on right). 
 Image conventions follow Fig.\ \ref{figimages}.\label{figroberts}}
\end{figure}

To make the analysis more tractable with the $I$ maps, we edge-enhanced
the shell structure while minimizing the background noise by processing
the maps with the Roberts' Cross operator \citep{roberts63}.
These maps are shown in Fig.\ \ref{figroberts}.
Because concentric arcs are azimuthally symmetric, the present
correlational method is unable to break the azimuthal degeneracy and
uniquely trace translational shifts of arc segments along the azimuthal
direction.  
This prevented us from following the differential proper-motion of
structures azimuthally away from the bipolar axis, and hence, from
independently confirming the bulk motion of the nebula and location of
the expansion center.  
Thus, for the present analysis we adopted the previously discovered
(1) bulk motion of the {\egg} at a rate of 
($13.7 \pm 2.0$, $10.2 \pm 2.0$) mas yr$^{-1}$, 
(2) the location of the illumination source/expansion center,
and 
(3) the distance to the {\egg} at $420 \pm 60$~pc
as found in our previous analysis \citep{ueta06}.

\begin{figure}
\begin{center}
\includegraphics[height=4.5in]{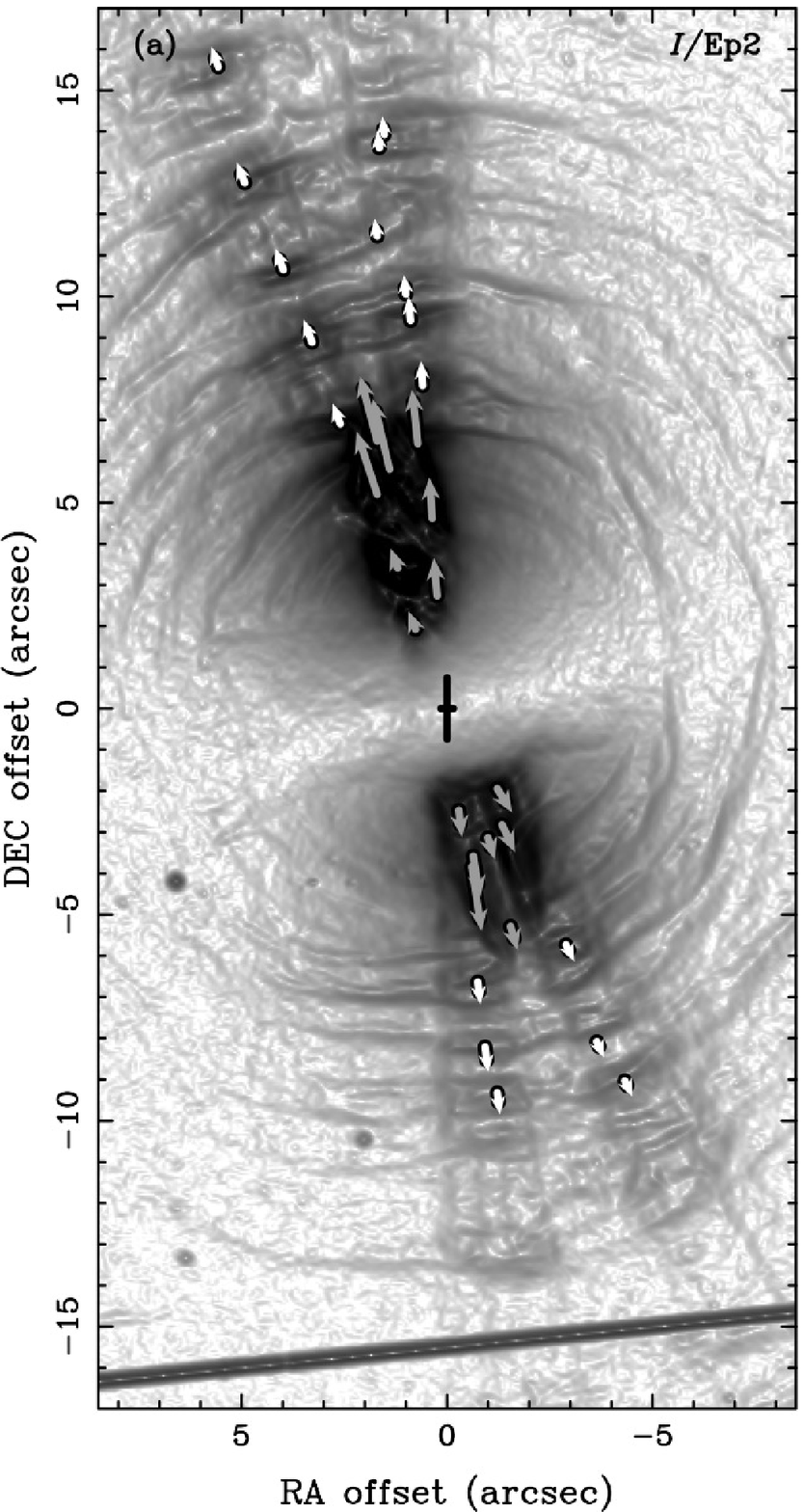}
\includegraphics[height=4.5in]{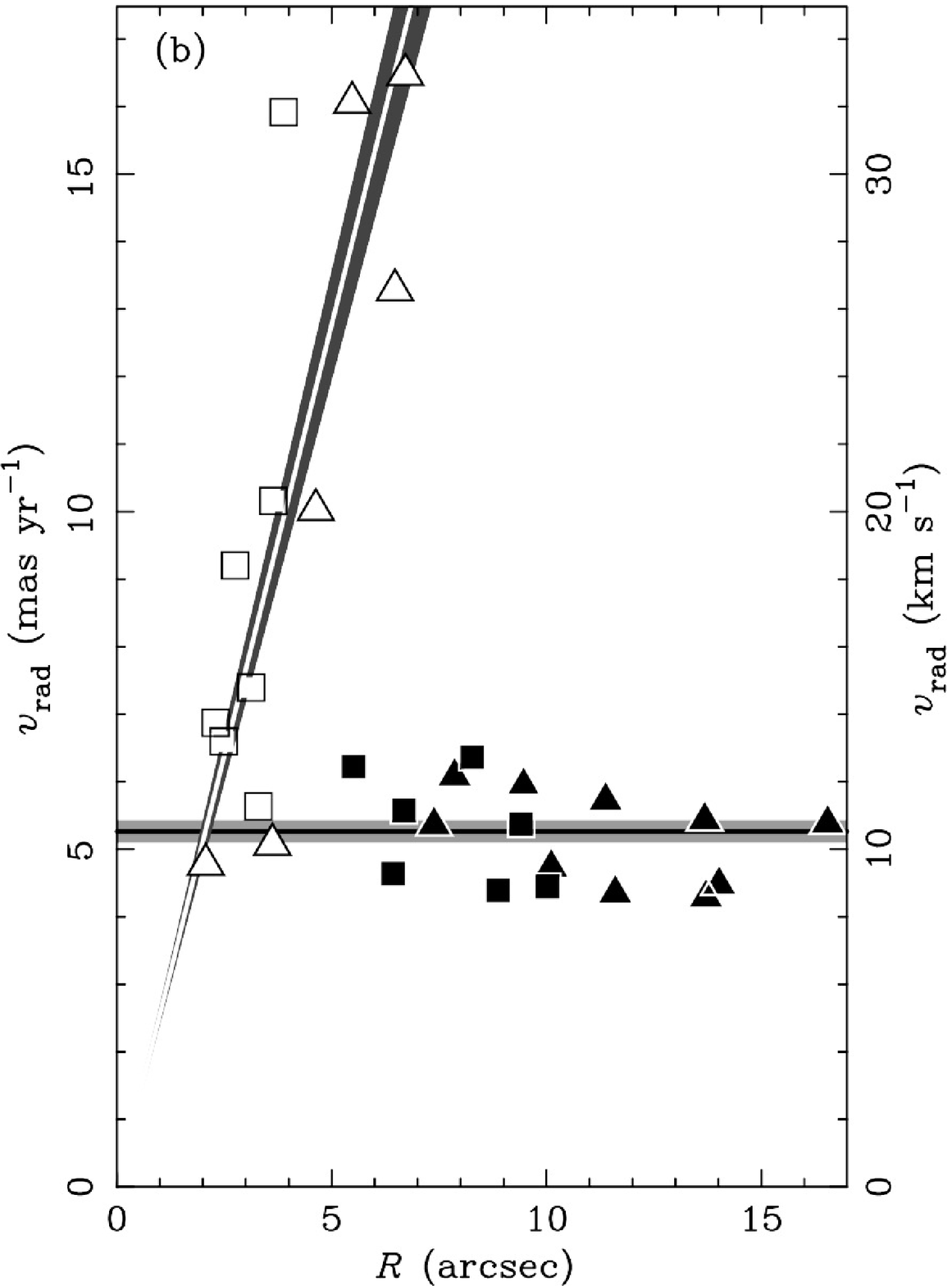} 
\end{center}
\caption{\label{figvectors}%
 (a) The edge-enhanced $I$ map (of epoch 2 in grayscale) of the 
 Egg Nebula overlaid with vectors of the projected radial component of
 the shell expansion.
 The length of the vectors depicts the relative velocity: gray vectors
 are of the lobes, while white vectors are of the arc segments.
 (b) The plot of $v_{\rm rad}$ (projected radial component of the
 differential proper-motion vector in arcsec~yr$^{-1}$
 and km~s$^{-1}$ assuming 420~pc) vs.\ $R$ 
(projected radial distance from the expansion center in arcsec) for
 these structures.   
Symbols indicate structures of
 the lobe (open triangle -- N lobe; open square -- S lobe) and arc
 segments (filled triangle -- arcs along the N searchlight beam;
 filled square -- arcs along the S searchlight beam)
The best-fit lines for the ensemble motion are shown as
 the white solid line with the gray zone of uncertainty (lobes) and
 black solid line with the light gray zone of uncertainty (arc segments).}
\end{figure}

We then selected 30 distinct local structures that register more than
five S/N with respect to the sky background distributed in the lobes 
(13 structures) and searchlight beams radially beyond the lobes (17
structures) and performed the correlational analysis to measure 
proper motion of these structures. 
To disregard the inadvertent azimuthal shifts, we extracted only the
projected radial component of the expansion of the structures
as presented in Fig.\ \ref{figvectors}a.

The same data are displayed as a plot of the projected radial component 
of the differential proper-motion vector ($v_{\rm rad}$ in mas~yr$^{-1}$
and km~s$^{-1}$) vs.\ projected radial distance from the location of the
illumination source/expansion center ($R$ in arcsec) in
Fig.~\ref{figvectors}b. 
For each of the lobes and arc segments, linear fitting was done to
characterize its proper motion. 
Upon fitting, we consider the following three plausible cases:
(1) a linear expansion (linear fitting of $v_{\rm rad}$ vs.\ $R$
anchored at the origin), 
(2) a constant-speed expansion (fitting by a horizontal line), and
(3) a more general linear expansion (linear fitting with a non-zero intercept),
and compared their Bayesian and Akaike information criteria (BIC and
AIC)\footnote{In statistics, BIC/AIC are measures of the relative
goodness of fit of a model which address issues of overfitting
(i.e., a model with a larger number of parameters usually fits better)
by penalizing models with a larger number of parameters.}
to identify the best-fit. 
Uncertainties in the projected speed and distances are roughly
5.0~mas~yr$^{-1}$ and 2$^{\prime\prime}$, respectively.
However, the {\sl relative} spacing among individual structures within
an ensemble remains the same and hence the slope of the linear fitting
is robust. 

The best-fit for the lobes yielded
$v_{\rm rad} = (2.56 \pm 0.15) \times R$~mas~yr$^{-1}$
or $(5.10 \pm 0.33) \times R$~km~s$^{-1}$ 
(the solid white line in the dark-gray zone), indicating that the lobes
are expanding linearly with the dynamical age of $390 \pm 25$~yr, which is 
roughly consistent with the previous results, $345 \pm 2$ and 
$266 \pm 40$~yr (\citealt{ueta06} and \citealt{balick12}, respectively).
\citet{cox00} measured the line of sight velocity of the tips of the CO
bipolar lobes (coinciding with the optical lobe tip pair F1/F2 in their
Figs.\ 2 and 3) as 
$6 \pm 1$~km~s$^{-1}$ at $6^{\prime\prime} \pm 1^{\prime\prime}$ away
from the center. 
By adopting the projected radial expansion velocity of $31 \pm 5$~km~s$^{-1}$
at the tip of the N lobe,
the inclination of the bipolar tips is computed as
$11^{\circ} \pm 2^{\circ}$ and the deprojected bipolar outflow velocity as
$32 \pm 5$~km~s$^{-1}$.
This outflow velocity is consistent with the medium-velocity component
found in CO \citep{young92,yamamura95}.

The best-fit for the arc segments along the searchlight beams beyond the
lobes turned out to be a constant-speed expansion at 
$v_{\rm rad} = (5.28 \pm 0.17)$~mas~yr$^{-1}$
or $10.52 \pm 0.34$~km~s$^{-1}$
(the solid black line in the light-gray zone), suggesting that 
the arcs have been ejected in an on-going basis and coasting away at a
constant speed, 
as opposed to being formed altogether at the same time and expanding
linearly.
This is a new finding from the present analysis with a larger field of
view with respect to the previous analysis by \citet{balick12}, in which
both growth patterns were found equally likely.
A detection of the central CO shell of $1^{\prime\prime}$ radius
expanding at around 10~km~s$^{-1}$ reported by \citet{cox00}, therefore,
may be related to the presence of these concentric arcs.
If so, the generation of the concentric arcs appears to have been
continuing even after the onset of the bipolar lobe expansion.

\subsubsection{Radial Expansion of the Concentric Arcs}

\begin{figure}
\begin{center}
 \includegraphics[width=\textwidth]{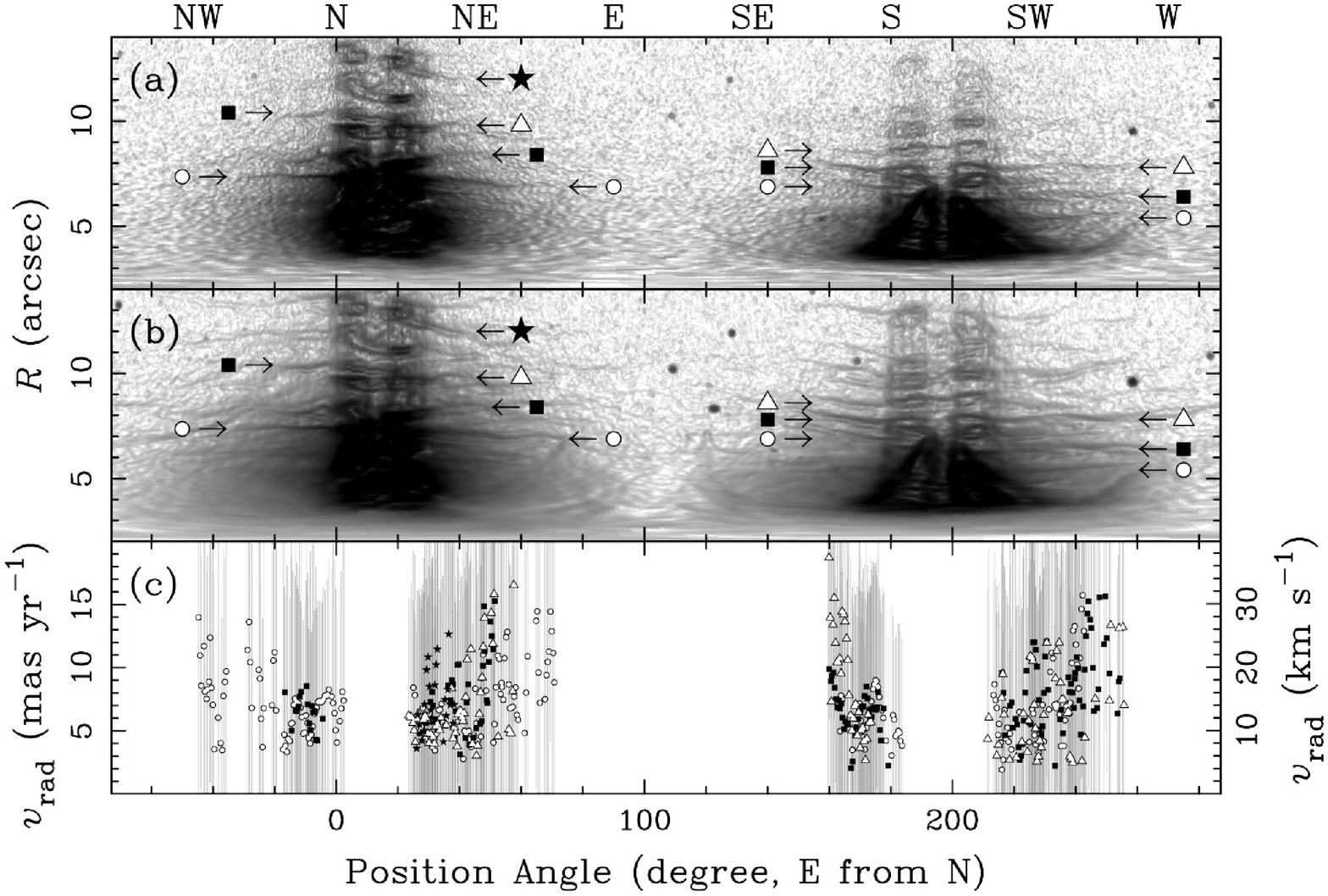}
\end{center}
\caption{The polar edge-enhanced Stokes $I$
 polar maps of the Egg Nebula for epoch 1 (top) and 
 epoch 2 (middle), with the distribution of measured projected radial
 expansion velocities in 12 most prominent arcs showing the apparent
 increasing trending of the radial velocity toward the equatorial
 plane (bottom). 
 The polar angle is the position angle measured from N to E, with the
 orientations indicated above the top frame.
 Each symbol in the bottom frame represents a differential radial-motion
 measurement for each 1$^{\circ}$ arc segment as a function of the
 position angle.
 The symbols also indicate the location of the arc in the polar maps.
 The expansion velocity, $v_{\rm rad}$, is given in mas~yr$^{-1}$ and 
 km~s$^{-1}$ (assuming 420~pc).
 Uncertainties of the radial velocity measurements are indicated by the 
 vertical lines attached to each symbol.
\label{figunrolled}}
\end{figure}

To further alleviate the azimuthal degeneracy issue of the concentric
arcs, we ``unrolled'' the edge-enhanced $I$ maps
with respect to the adopted location of the expansion center to make 
polar ($r$, $\Theta$) maps, where $r$ and $\Theta$ are the radial
distance from the center and the position angle measured from N,
respectively.
Then, we traced the projected radial motion of concentric arcs 
as a function of the position angle
by performing the same correlational analysis only along the radial
direction for each $1^{\circ}$ azimuthal arc segment as long as these
arc segments register more than three-$\sigma$ of the background
(Fig.~\ref{figunrolled}a,b).

Fig.~\ref{figunrolled}c shows the distribution of the projected
radial velocities of these 1$^{\circ}$ segments measured along 12 most
prominent arcs.
While uncertainties are relatively large owing to a large S/N
discrepancy between the two-epoch maps, the present data suggest
that the projected radial velocities tend to
(1) converge to a minimum of about 10~km~s$^{-1}$ toward the
regions of the searchlight beams/bipolar axis, and 
(2) increase by a factor of about a few as moving toward the equatorial
plane.
While the scatterer in the measured radial velocities is large, 
the increasing tendency appears genuine.

If there were a well-defined azimuthal/latitudinal gradient in the
radial expansion velocity field, it should have manifested itself in 
the overall structure of the concentric arcs -- the arcs should have
been elongated along the equatorial plane at least by a factor of a few.
However, the arcs strikingly concentric.
Therefore, it is
most likely that there is a faster outflow ($> 20 - 30$~km~s$^{-1}$) 
in spatially restricted regions along the equatorial plane and that
arc segments along the equatorial plane must have been blown out by such
an outflow -- the higher-velocity segments near the end of the
observed arcs are most likely the arc edges being torn off
by the suspected outflow.


Combined altogether, it appears that there are three distinct outflow
components in the circumstellar environs of the {\egg}.
There is a generally spherically symmetric steady outflow with a
periodical modulation at about 10~km~s$^{-1}$, which is responsible for
the concentric arcs. 
Based on the arc segments seen farthest from the center in the
higher-S/N epoch 2 map, this periodic outflow appears to have continued
at least for the past $\sim 5,500$~yr at the interval of $50 - 400$~yr
at 420~pc. 

This generally symmetric shell structure seems to be disrupted by two
distinct outflow components, one along the bipolar axis and the other 
along the equatorial plane.
It has already been shown that the bipolar lobes have been being
excavated by a linear expansion for the past $300 - 400$~yr with the tip
velocity being about 30~km~s$^{-1}$. 
Circumstantially, this linearly expanding outflow along the polar axis
seems to have punched holes out of the central dust cocoon,
through which most of the dust-scattered light streams out along the
bipolar axis to illuminate the lobes.
The present analysis of the concentric arcs has indicated the presence
of another outflow component along the equatorial plane: its velocity
appears to be greater than $20 - 30$~km~s$^{-1}$ based on 
the ways that concentric arcs are being disrupted along the equatorial
plane.
Other observational details of this outflow including the mode and
duration, however, remain unclear.

\subsection{Origins of the Outflow Components}

The apparent superposition of three outflow components in
the {\egg} is dynamically very intriguing.
While theoretical consideration of the latitudinal dependence of the 
{\sl velocity} field has been rare (e.g., \citealt{dwarkadas96}),
in studying the influence of the central binary system to
the AGB wind structure, \citet{mm99} numerically discovered that 
(1) the latitudinal dependence of the gas density is always accompanied
by a latitudinal decline in the outflow velocity field, and 
(2) there can be a positive correlation between the latitudinal density
and velocity contrasts (i.e., outflow velocity becomes faster where
density is greater; their model 4),
contrary to many dust-driven models that {\sl assume} a velocity
profile that increases toward the polar direction.

This means that the present case may provide rare observational
evidence for such a colatitudinally-dependent outflow velocity field
resulting from the egg-beater mass-loss modulation by the central binary
system.
While spiral modulation of mass-loss ejecta due to the central binary
system has been seen around AGB stars for near pole-on cases 
(AFGL 3068; \citealt{mh06},  
Mira; \citealt{mayer11}, 
R Scl; \citealt{m12}),
the present data may lend support for an edge-on case in which 
(1) the spiraling trails of mass-loss ejecta appear as concentric arcs, and 
(2) the colatitudinal velocity field promotes an equatorial outflow
fast enough to disrupt the ejecta (e.g., \citealt{mm99}).

One issue in the present analysis is that the discrepancy in the signal
strengths in the two-epoch data set (due to much lesser S/N of epoch 1
data) resulted in rather large uncertainties in proper-motion
measurements.  
At any rate, both theoretical and observational work is scarce at best
to establish the connection between the concentric arcs and the binary 
modulation of stellar winds.
Therefore, another epoch of {\hst} observations to obtain data with a
comparable S/N to the epoch 2 data are extremely interesting in 
quantifying the colatitudinal dependence of the radial velocity field to
be compared against the latest theoretical expectations 
(e.g., \citealt{mp11,kt12}).
Also, mapping these concentric arcs in CO at the comparable spatial
resolution of the full {\sl ALMA} is worthwhile (e.g., \citealt{m12}) to
do understand kinematics between the gas and dust components of these outflows.

\section{Concluding Summary}

We performed differential proper-motion and radial-motion analyses
on both the signature bipolar lobes and concentric arcs of the {\egg}
using two-epoch optical data taken by the {\hst}.
Our method, based on 
(1) aligning two-epoch images using stationary background stars as
spatial anchors, and
(2) measuring the amount translational movement of specific local
structures by a correlational method, 
is more robust than the previous analysis by \citet{balick12} that
assumed symmetry in the shell expansion.
Our findings are that;
 \begin{enumerate}
  \item the {\egg} is illuminated optically by a two-step mechanism, in
	which the direct starlight is first channeled into the bipolar
	cavities and then scattered off to the rest of the nebula,
	
  \item there is a ``dustsphere'' around the central star
	($1.0 \times 10^3$~AU by 
	$1.6 \times 10^3$ AU at 420~pc), most likely representing the
	surface of the central dust concentration at which the
	line-of-sight optical depth in the $V$ band becomes significantly
	larger than unity, 
	
  \item the bipolar lobes expand linearly for the past $390 \pm 25$~yr,
	excavating into an otherwise symmetric concentric shells 
	as found by the previous analyses \citep{ueta06,balick12},

  \item	arc segments found along the searchlight beams beyond the
	bipolar lobes all move at similar projected radial velocity 
	at $10.52 \pm 0.34$~km~s$^{-1}$ for the past $\sim 5,500$~yr,

  \item the projected radial expansion velocity field of the concentric
	arcs shows an increasing trending toward the equatorial plane by
	a factor of a few, representing torn-off edges of the arcs
	disrupted by a fast equatorial outflow at $>20-30$~km~s$^{-1}$, 
	and

  \item the present case may provide observational evidence for the
	binary modulation of stellar winds, as predicted by \citet{mm99},
	which generates	
	the colatitudinal dependence in the radial expansion velocity
	field and the spiral-shocks of the mass-loss ejecta appearing
	concentric arcs in near edge-on orientations.
 \end{enumerate}

 \acknowledgments

 Support for this work was provided by NASA/STScI through a grant 
 HST-AR-12157.01-A and 
 by the University of Denver through a Professional Research
 Opportunities for Faculty (PROF) grant.
 Authors thank Dr.\ D.\ A.\ Weintraub for explaining his method to
 pinpoint the location of the illumination source in the imaging
 polarimetry data thoroughly to us.
 Authors also acknowledge Drs.\ Robert E.\ Stencel
 and Djazia Ladjal for their excellent comments which improved the work. 
 
{\it Facilities:} \facility{HST (WFPC2,ACS/WFC)}.



\begin{deluxetable}{llcccc}
\tabletypesize{\normalsize}
\tablecaption{Two-Epoch Imaging-Polarimetry Observing Log of the Egg Nebula\label{obs}}
\tablewidth{0pt}
\tablehead{%
\colhead{} & \colhead{} & \colhead{} & 
\colhead{EXP TIME} & \colhead{} & \colhead{} \\
\colhead{DATE} & \colhead{INSTRUMENT} & \colhead{FILTER} & 
\colhead{(sec)} & \colhead{PID} & \colhead{DATASET ID}}
\startdata
1995 Jul 17 (epoch1) & WFPC2   & F606W/POLQ & 300.0 & 6221 &
 U2RC0301/2/3T \\
2002 Oct 16 (epoch2) & ACS/WFC & F606W/POLV & 999.0 & 9586 &
 J8GH5501/2/31 
\enddata
\end{deluxetable}


\begin{deluxetable}{ccc}
\tabletypesize{\normalsize}
\tablecaption{Two-Epoch Aperture Photometry\label{photom}}
\tablewidth{0pt}
\tablehead{
\colhead{} & \colhead{} & \colhead{$F_{\lambda}$} \\
\colhead{Date} & \colhead{Aperture} & 
\colhead{($10^{-12}$ erg cm$^{-2}$ s$^{-1}$ \AA$^{-1}$)} }
\startdata
1995 Jul 17 & All    & $1.85 \pm 0.02$ \\
            & N Lobe & $1.38 \pm 0.01$ \\
            & S Lobe & $0.47 \pm 0.01$ \\
2002 Oct 16 & All    & $1.18 \pm 0.01$ \\
            & N Lobe & $0.90 \pm 0.01$ \\
            & S Lobe & $0.27 \pm 0.01$ 
\enddata
\end{deluxetable}



\begin{thebibliography}{}
  \bibitem[Balick et al.(2012)]{balick12}
		  Balick, B., Gomez, T., Vinkovi\'{c}, D., Alcolea. J.,
		  Corradi, R.\ L.\ M., \& Frank, A.
		  2012, \apj, 745, 188

  \bibitem[Biretta \& McMaster(1997)]{bm97}
		  Biretta, J., \& McMaster, M.
		  1997, Instrument Science Report WFPC2 97-11
		  (Baltimore: STScI)

  \bibitem[Cox et al.(2000)]{cox00}
		  Cox, P., Lucas, R., Huggins, P.\ J., Forveille, T.,
		  Bachiller, R., Guilloteau, S., 
		  Maillard, J.\ P., \& Omont, A. 
		  2000, \aap, 353, L25

  \bibitem[Currie et al.(1996)]{currie96}
		  Currie, D.\ G., Dowling, D.\ M., \& Shaya, E.\ J., et al. 
		  1996, \aj, 112, 1115

  \bibitem[Dwarkadas, Chevalier, \& Blondin(1996)]{dwarkadas96}	
		  Dwarkadas, V.\ V., Chevalier, R.\ A., \& Blondin, J.\ M.
		  1996, \apj, 457, 773

  \bibitem[Fruchter et al.(2009)]{md}
		  Fruchter, A., \& Sosey, M., et al. 
		  2009, The MultiDrizzle Handbook, Ver.\ 3.0, (Baltimore: STScI)

  \bibitem[Gonzaga et al.(2010)]{gonzaga10}
		  Gonzaga, S., \& Biretta, J., et al. 
		  2010, WFPC2 Data Handbook, Ver.\ 5.0 (Baltimore: STScI)

  \bibitem[Gonzaga et al.(2011)]{gonzaga11}
		  Gonzaga, S., et al., 
		  2011, ACS Data Handbook, Ver.\ 6.0 (Baltimore: STScI)

  \bibitem[Goto et al.(2002)]{goto02}
		  Goto, M., Kobayashi, N., Terada, H., \& Tokunaga, A.\ T.
		  2002, \apj, 572, 276

  \bibitem[Hrivnak et al.(2001)]{hrivnak01}
		  Hrivnak, B.\ J., Kwok, S.\ \& Su, K.\ Y.\ L.
		  2001, \aj, 121, 2775
	
  \bibitem[Hrivnak et al.(2010)]{hrivnak10} 
		  Hrivnak, B.\ J., Lu, W., Maupin, R.\ E., \& Spitzbart, B. D.
		  2010, \apj, 709, 1042
	  
  \bibitem[Kim \& Taam(2012)]{kt12}
		  Kim, H., \& Taam, R.\ E.
		  2012, \apj, 759, 59

  \bibitem[Kwok(1993)]{kwok93}
		  Kwok, S.
		  1993, \araa, 31, 63

  \bibitem[Latter et al.(1993)]{latter93}
		  Latter, W.\ B., Hora, J.\ L., Kelly, D.\ M., Deutsch,
		  L.\ K., \& Maloney, P.\ R. 
		  1993, \aj, 106, 260

  \bibitem[Maercker et al.(2012)]{m12}
		  Maercker, M., Mohamed, S., Vlemmings, W.\ H.\ T.,
		  Ramstedt, S., Groenewegen, M.\ A.\ T., Humphreys, E.,
		  Kerschbaum, F., Lindqvist, M., Olofsson, H., Paladini, C.,
		  Wittkowski, M., de Gregorio-Monsalvo, I., \& Nyman, L.-A.
		  2012, \nat, 490, 7419
		  
  \bibitem[Mauron \& Huggins(2006)]{mh06}
		  Mauron, N., \& Huggins, P.\ J.
		  2006, \aap, 452, 257

  \bibitem[Mayer et al.(2011)]{mayer11}
		  Mayer, A., Jorissen, A., Kerschbaum, F., Mohamed, S., van Eck,
		  S., Ottensamer, R., Blommaert, J.\ A.\ D.\ L., Decin, L.,
		  Groenewegen, M.\ A.\ T., Posch, Th., Vandenbussche, B., \& Waelkens, C.
		  2011, \aap, 531, L4

  \bibitem[Mohamed \& Podsiadlowski(2011)]{mp11}
		  Mohamed, S., \& Podsiadlowski, Ph.
		  2011, in ASP Conf.\ Ser.\ 445, 
		  Why Galaxies Care about AGB Stars II: Shining Examples
		  and Common Inhabitants,
		  eds.\ F.\ Kerschbaum et al.\ (San Francisco: ASP), 355

  \bibitem[Mastrodemos \& Morris(1999)]{mm99}
		  Mastrodemos, N., \& Morris, M.
		  1999, \apj, 523, 357

  \bibitem[Morse et al.(2001)]{morse01}
		  Morse, J.\ A., Kellogg, J.\ R., Bally, J., Davidson,
		  K., Balick, B., \& Ebbets, D. 
		  2001, \apj, 548, L207

  \bibitem[Ney et al.(1975)]{ney75}
		  Ney, E.\ P., Merrill, K.\ M., Becklin, E.\ E., Neugebauer, G., \&
		  Wynn-Williams, C.\ G. 
		  1975, \apj, 198, L129
	
  \bibitem[Oke(1974)]{oke74}
		  Oke, J.\ B.
		  1974, \apjs, 27, 21

  \bibitem[Roberts(1963)]{roberts63}
		  Roberts, L.\ G.
		  1963, Lincoln Laboratory Technical Report, 315,
		  22 (Bostin: MIT)

  \bibitem[Sahai et al.(1998a)]{sahai98b}
		  Sahai, R., Hines, D.\ C., Kastner, J.\ H., et al.
		  1998, \apjl, 492, 163

  \bibitem[Sahai et al.(1998b)]{sahai98}
		  Sahai, R., Trauger, J.\ T., Watson, A.\ M., et al. 
		  1998, \apj, 493, 301

  \bibitem[Su(2004)]{su04}
		  Su, K.\ Y.\ L.
		  2004, in ASP Conf.\ Ser.\ 313, Asymmetrical Planetary Nebulae III,
		  eds.\ M.\ Meixner et al.\ (San Francisco: ASP), 247

  \bibitem[Ueta et al.(2005)]{ueta05}
		  Ueta, T., Murakawa, K., \& Meixner, M.
		  2005, \aj, 129, 1625

  \bibitem[Ueta et al.(2006)]{ueta06}
		  Ueta, T., Murakawa, K., \& Meixner, M.
		  2006, \apj, 641, 1113

  \bibitem[Ueta et al.(2007)]{ueta07}
		  Ueta, T., Murakawa, K., \& Meixner, M.
		  2007, \aj, 133, 1345

  \bibitem[Ueta et al.(2012)]{ueta12}
		  Ueta, T., Murakawa, K., \& Meixner, M.
		  2012, \apj, 756, 103

  \bibitem[van Dokkum(2001)]{vd01}
		  van Dokkum, P.\ G.
		  2001, \pasp, 113, 1420

  \bibitem[Van Winckel(2003)]{vanwinckel03}
		  Van Winckel, H.
		  2003, \araa, 41, 391

  \bibitem[Weintraub \& Kastner(1993)]{wk93}
		  Weintraub, D.\ A., \& Kastner, J.\ H.
		  1993, \apj, 411, 767

  \bibitem[Weintraub et al.(2000)]{weintraub00}
		  Weintraub, D.\ A., Kastner, J.\ H., Hines, D.\ C., \&
		  Sahai, R.
		  2000, \apj, 531, 401

  \bibitem[Yamamura et al.(1995)]{yamamura95}
		  Yamamura, I., Onaka, T., Kamijo, F., Deguchi, S., \& Ukita, N.
		  1995, \apj, 439, L13

  \bibitem[Young et al.(1992)]{young92}
		  Young, K., Serabyn, G., Phillips, T.\ G., Knapp, G.\ R.,
		  G\"{u}sten, R., \& Schulz, A.  
		  1992, \apj, 385, 265

 \end{thebibliography}
\end{document}